\documentclass[11pt,a4paper,amssymb,amsmath,amsbsy]{article}
\usepackage{graphicx}
\usepackage{slashed}
\usepackage{bbold}
\usepackage{float}
\usepackage{amsfonts}
\usepackage{subcaption}
\usepackage{amsmath}
\usepackage{amssymb}
\usepackage{enumerate}
\usepackage{latexsym}
\usepackage{color}
\usepackage{pifont}
\usepackage{tikz}
\usepackage{colortbl}
\usepackage{fancybox}
\usepackage{epstopdf}
\usepackage{hyperref}
\usepackage{latexsym}
\usepackage{fancybox}
\usepackage{tikz}
\usepackage{colortbl}
\usepackage[T1]{fontenc}
\usepackage[utf8]{inputenc}
\usepackage{graphicx}
\usepackage{braket}
\providecommand{\keywords}[1]{\textbf{\textbf{keywords:}} #1}
\usepackage[symbol]{footmisc}
\begin{document}

\thispagestyle{empty}

\begin{center}
\vspace{1.8cm}

{\Large \textbf{Quantifying quantum correlations in noisy Gaussian channels.}}


\vspace{1.5cm} \textbf{Y. Lahlou}$^{a,}$\footnote{Corresponding author.}\footnote{%
email: \textsf{youness\_lahlou@um5.ac.ma}}, \textbf{L. Bakmou}$^{a,}${\footnote{%
email: \textsf{baqmou@gmail.com}}} \textbf{B. Maroufi}$^{b,}${\footnote{%
email: \textsf{mbouchra@hotmail.com}}} and \textbf{M. Daoud}$^{b,}${%
\footnote{%
email: \textsf{m\_daoud@hotmail.com}}}

\vspace{0.5cm}

$^{a}$\textit{LPHE-MS, Department of Physics, Faculty of Sciences, Mohammed V University, Rabat, Morocco}\\[0.5em]
$^{b}$\textit{Department of Physics, Faculty of Sciences, Ibn Tofail
University, K\'{e}nitra, Morocco}

\vspace{3cm} \textbf{Abstract}
\end{center}

\baselineskip=18pt

The Gaussian states are essential ingredients in many tasks of quantum information processing. The presence of the noises imposes limitations on achieving these quantum protocols. Therefore, examining the evolution of quantum entanglement and quantum correlations under the coherence of Gaussian states in noisy channels is of paramount importance. In this paper,  we propose and analyze a scheme that aims to specify and examine the dynamic evolution of the quantum correlations in two-modes Gaussian states submitted to the influence of the Gaussian thermal environment. We describe the time evolution of the quantum correlations in an open system consisting of two coupled bosonic modes embedded in a Gaussian thermal environment. We discuss the influence of the environment in terms of the initial parameters of the input states. The quantum correlations are quantified using  Gaussian interferometric power and the Gaussian entanglement of formation. The behavior of these quantum correlations quantifiers is strictly dependent on the parameters of the input states that are employed. We show that the Gaussian interferometric power is a measurement quantifier that can capture the essential quantum correlations beyond quantum entanglement. In addition, we show that the Gaussian interferometric power is less influenced than the Gaussian entanglement of formation.

\vspace{0.7cm}
\newpage
\section{Introduction\label{sec1}}
Quantum information aims to manipulate and transmit information by exploiting the quantum features, in which all pieces of information are symbolized and encoded in quantum states. Recently, the development of the quantum information processing is experiencing an incredible increase, which is motivated by quantum improvement communication \cite{1,2}, quantum measurement \cite{3}, and quantum computing \cite{4}. Indeed, most of the quantum protocols were proposed initially for quantum-discrete variables (DV)\cite{5}, especially quantum bits using physical systems such as photons, ions, atoms, solid-state, superconducting devices, and nuclear magnetic resonance \cite{6}. More recently, considerable attention has been devoted to examining the use of continuous variable (CV) systems in implementing quantum information processing protocols \cite{7,8,lahlou2019quantum,9}. In order to perform the different tasks of quantum information processes,  the CV systems are easy to manipulate in labs than the manipulation of the quantum bit.  For this reason, many works have been devoted extensively to CV systems from theoretical as well as experimental points of view.

Gaussian states, for instance, squeezed coherent beams and twin beams \cite{10},  are a particular family of CV systems that can generate and manipulate experimentally\cite{werner2001bound, diguglielmo2011experimental}. In addition,  the Gaussian states have a specific elegant formalism and provide beneficial tools to encode and process quantum information with continuous variables due to the limitation of the degrees of freedom, which is limited only to the displacement vector and the covariance matrix. As a matter of fact, the Gaussianity of quantum states is preserved during many quantum transformations operations, either unitary \cite{schumaker1986quantum, olivares2012quantum} or non-unitary \cite{wolf2007quantum, bellman1957markovian, olivares2003optimized, Bakmou}. These types of quantum states have been used successfully to carry out quantum teleportation and quantum error correction \cite{banchi2015quantum, ralph2011quantum}. Besides, Gaussian states were exploited extensively in manipulating quantum cloning entanglement purification in the test of quantum non-locality \cite{nha2004proposed, wiseman2007steering}.

Unquestionably,  the different kinds of quantum correlations \cite{12}, like quantum entanglement \cite{13,14}, non-locality, are playing a paramount role in various areas of quantum information processing. Indeed, the entanglement was recognized as an essential resource employed for quantum computing \cite{15},  cryptographic protocols \cite{16}, and realized quantum teleportation \cite{12}. In addition, quantum entanglement has been used to improve the detection in Gaussian quantum metrology \cite{nichols2018multiparameter,bakmou2020multiparameter} and quantum measurement strategy \cite{st}.  Recently, quantum entanglement, in CV specifically in Gaussian states, has been proved as a valuable tool for improving optical resolution \cite{18,lahlou2020gaussian}, spectroscopy \cite{spec}, interferometry \cite{girolami2014quantum}, tomography \cite{tomography}, and discrimination of quantum operations \cite{dis}.

The implementation of quantum information protocols requires the transfer of the quantum correlations under the physical non-unitary channels. In these types of physical channels, the implementation tasks of the quantum information process are a challenging procedure due to the unavoidable interaction with the environment. This leads to the degradation of quantum correlations due to decoherence and dissipation generated by losses and noise \cite{no}. Indeed in the last decades, many works were dedicated to realistic descriptions of quantum decoherence and the dynamics of quantum correlations in open quantum systems, particularly in CV systems.  Among these works, let us mention those based on the study of the dynamics of quantum correlations in colored-noise environments \cite{cl}, system-reservoir \cite{re}, two qubits coupled to bosonic baths \cite{bo}, and the dynamics of quantum correlations under decoherence \cite{de}. In this context, the main objective of our work is to continue in the same direction of investigating the evolution of quantum correlations in noisy Gaussian channels.

In this paper, we examine the dynamics of quantum correlations in a system composed of two bosonic modes immersed in a  Gaussian thermal reservoir, such that each one interacts with its thermal bath. The paper is structured as follows. Section \ref{sec2} reviews the basic concept of Gaussian states that are required to realize our purpose, and we present the basic concepts of Gaussian noises channels.  In the next section \ref{sec3}, we devoted ourselves to a declaration of the quantifiers used to study the dynamics of quantum correlations in the proposed protocol. Section \ref{sec4} is dedicated to discussing and analyzing the results realized in dealing with the dynamics of quantum correlations in the noises environment channels by using the different quantum quantification measures. We end our paper with a conclusion.

\section{ PRELIMINARIES OF GAUSSIAN STATES \label{sec2}}
To start,  we consider $m$-bosonic modes. Each mode $k=1, 2,...,m$  is described by the annihilation and creation operators ${{\hat a}_k}$ and $\hat a_k^\dag$, that satisfies the commutation relations. $\left[ {{{\hat a}_k},\hat a_l^\dag } \right] = {\delta _{kl}}$. The Hilbert space of the system is the tensor product of the infinite dimensional Fock space $\mathcal{H }= \mathop  \otimes \limits_{k = 1}^m {\mathcal{F}_k}$, where, the Fock space associared with the mode $k$ is characterized by the number basis $\left\{ {{{\left| m \right\rangle }_k}} \right\}_{k = 1}^m$, spanned by the vectors of this basic are eigenstates of the number operator $\hat a_k^\dag {\hat a_k}$. The free Hamiltonian of non-interacting modes is given by the Hamiltonian $H = \sum\limits_{k=1}^m {\left( {\hat a_k^\dag {{\hat a}_k} + \frac{1}{2}} \right)}$.  The corresponding quadrature operators (position and conjugate momenta) for the $k^{\text{th}}$ mode are
\begin{equation}
{Q_k} = \hat a_k^\dag  + {{\hat a}_k}, \hspace{1cm} {P_k} = i\left( {\hat a_k^\dag  - \hat a} \right).
\end{equation}
Here we set $\hbar=2$.  The commutation relations between the quadratics operators write $\left[ {{Q_k},{P_l}} \right] = 2i{\delta _{kl}}$. These structure relations can be cast in compact form, using the blocks vector of quadratics operators  $\mathbf{\hat R} = {\left( {\begin{array}{*{20}{c}}
		{{\hat Q_1},}&{{\hat P_1},}&.&.&.&,&{{\hat Q_m},}&{{\hat P_m},}
		\end{array}} \right)^T}$. In this compact form,  the vector of quadratics operators satisfies the following commutation relation 
\begin{equation}
\left[ {{{\hat R}_k},{{\hat R}_l}} \right] = 2i{\Omega _{kl}}, 
\end{equation}
where $\Omega _{kl}$ are the elements of the $2m \times 2m$ symplectic matrix defined by
\begin{equation}
\Omega  = \mathop  \oplus \limits_{k = 1}^m \omega, \hspace{1cm}\omega  = \left[ {\begin{array}{*{20}{c}}
	0&1\\
	{ - 1}&0
	\end{array}} \right].
\end{equation}
For a given $m$-mode bosonic system, the density operator has an equivalent representation in terms of the quasi-probability distribution defined in phase space, namely the characteristic function or the Wigner function \cite{vogel, tomography}. This representation called Wigner representation \cite{tatarski}, such that, all the information of the quantum system encoded in the density operator $\rho$, are analogically encoded in the displacement vector $\boldsymbol{d}= \boldsymbol{\left\langle {\hat R} \right\rangle }$ and the covariance matrix that defined in the phase space by
\begin{equation}
{\sigma _{kl}} \equiv {[{\bf{\sigma }}]_{kl}} = \frac{1}{2}\left\langle {\left\{ {{\hat{R}_k},{\hat{R}_l}} \right\}} \right\rangle  - \left\langle {{\hat{R}_l}} \right\rangle \left\langle {{\hat{R}_k}} \right\rangle, 
\end{equation}
where $\left\{ {A,B} \right\}=AB+BA$ denotes the anti-commutator and $\left\langle {\hat{O}} \right\rangle=Tr\left[{\rho}\hspace{0.1cm}\hat{O}\right]$ is the expectation value of the operator $\hat{O}$. The covariance matrix $\sigma$ is a $2m \times 2m$ real symmetric matrix which is defined positive and satisfy the following uncertainty relation \cite{simon1987gaussian}
\begin{equation}
\sigma+i \Omega \geq 0. \label{Eq. 5}
\end{equation}
From the diagonal elements of Eq. (\ref{Eq. 5}), one can easily derive the usual
	Heisenberg relation for position and momentum, i.e
	\begin{equation}
		\mathtt{Var}\left({\hat Q}_i\right)\mathtt{Var}\left({\hat P}_i\right)\ge 1.
\end{equation}
 The description of the formalism of Gaussian states is quite simple, and a large class of transformations acting on this kind of state is also simple to describe. In general, a quantum state that undergoes a quantum operation is described by a map ${\cal E}: \rho  \to {\cal E}(\rho )$, which is completely positive. In fact, we can distinguish between two types of maps; the first is completely positive and trace-preserving i.e. $Tr\left[ {{\cal E}( \rho )} \right] = 1$,  the second is a linear map that is completely positive and trace decreasing i.e. $Tr\left[ {{\cal E}( \rho )} \right] \le 1$. The first category is associated with closed quantum channels which admit reversible maps. These are described by unitary transformations, $\hat U{\hat U^\dag } = \mathbb{1}$, that transform a quantum state according to 
\begin{equation}
\hat U:\rho  \to \hat U \rho {\hat U^\dag }. \label{Eq. 6}
\end{equation}
The unitary transformations that preserve the Gaussian characteristic of a given quantum state are called Gaussian unitary channels. They are generated by the operation $\hat U = \exp ( - i\hat H)$, where $\hat{H}$ is the Hamiltonian of the system having forms that are second-order polynomials in quadrature operators. These Gaussian unitary transformations are described, in the phase space, by the symplectic transformations, which are linear transformations preserving the symplectic form 
\begin{equation}
\hat S \Omega \hat S^{T}=\Omega,
\end{equation}
where $\hat S$ is a $2m\times2m$ real symplectic matrix action on the covariance matrix.  In the phase space, the projection of the rule of Eq. (\ref{Eq. 6}) gives 
\begin{equation}
\hat S:\hat \rho  \to \hat S \sigma {\hat S^\dag }. 
\end{equation}
Besides the Gaussian unitary channels associated with the first type of maps ${\cal E}: \rho  \to {\cal E}( \rho )$, the Gaussian no-unitary channels (Noisy Gaussian channels) forming in the second type of maps are of unusual importance in the implementation of quantum information processing.

   Noisy Gaussian channels or Gaussian no-unitary channels are generally subject to noise and loss of quantum coherence due to interaction with the environment. As matter of fact, the dynamics of the $m$-mode bosonic system evolving under the effect of the noisy channel can be described by the following master equation:
\begin{equation}
\dot{\rho}=\frac{\Gamma}{2}\left\{(N+1) \mathcal{L}[\hat a]+N \mathcal{L}\left[\hat a^{\dagger}\right]-M^{*} \mathcal{D}[\hat a]-M \mathcal{D}\left[\hat a^{\dagger}\right]\right\} \rho, \label{equ1}
\end{equation}
where $N \in \mathbb{R}$ and $M \in \mathbb{C}$ represent the effective number of photons and the squeezing parameter of the bath respectively, while $\Gamma$ is the overall damping rate, $\mathcal{L}[\hat O] \rho=2 \hat O \rho \hat O^{\dagger}-\hat O^{\dagger} \hat O \rho-\rho \hat O^{\dagger} \hat O$ and $\mathcal{D}[\hat O] \rho=2 \hat O \rho \hat O-\hat O \hat O \rho-\rho \hat O \hat O$ are Lindblad super-operators. The expressions proportional to $\mathcal{L}[\hat a]$ and  $\mathcal{L}\left[\hat a^{\dagger}\right]$ denote the losses and linear, phase-insensitive, amplification processes, respectively, while the expressions proportional to $\mathcal{D}[\hat{a}]$ and $\mathcal{D}\left[\hat{a}^{\dagger}\right]$ denote the phase-dependent fluctuations. The positivity of the density matrix imposes the constraint $|M|^{2} \leq N(N+1)$.
When arriving at thermal equilibrium, i.e. for $M=0, N$ synchronize with the average number of thermal photons.\\
 Now, we consider the $m$-mode input Gaussian state $\rho_{inp}$ with the second moment $\sigma_{{\rm{inp }}}$. The non-unitary evolution preserves the Gaussian characteristic and the time dynamics action of the dissipation on the second moments at a time $t$ is described by the following equations of motion \cite{ferraro}
\begin{equation}
\dot{\boldsymbol{\sigma}}=-\Gamma\left(\boldsymbol{\sigma}-\boldsymbol{\sigma}_{\infty}\right)
\end{equation}
which yields to the following solution
\begin{equation}
\boldsymbol{\sigma}(t)=e^{-\Gamma t} \boldsymbol\sigma_{{\rm{inp }}}+\left(1-e^{-\Gamma t}\right) \boldsymbol{\sigma}_{\infty}\label{sigma t}
\end{equation}
 with 
${\sigma _\infty } = \mathop  \oplus \limits_{k = 1}^m {\sigma _{k,\infty }}$ is an asymptotic diffusion covariance matrix related to the effect imposed by the Gaussian environment. It is given only in terms of the bath parameters as 
\begin{equation}
\boldsymbol{\sigma}_{k, \infty}= \frac{1}{2}\left(\begin{array}{cc}
\left(\frac{1}{2}+N\right)+\Re \mathrm{e}[M] & \Im \mathrm{m}[M] \\
\Im \mathrm{m}[M] & \left(\frac{1}{2}+N\right)-\Re \mathrm{e}[M]
\end{array}\right).
\end{equation}
This diffusion matrix represents the asymptotic covariance matrix when the initial state is Gaussian.
\section{ Dynamics of  Gaussian interferometric power and entanglement of formation \label{sec3}}
\subsection{ Gaussian interferometric power}
The concept of Gaussian interferometric power (GIP)  has been introduced in \cite{adesoo}. It provides us with another powerful tool to quantify the ability of two-mode Gaussian to estimate the phase shift parameters in the metrology schema known, in the literature, as black-box interferometry. In this scheme,  for a bosonic continuous variable system of modes A and B, one consider $\rho_{AB}$ as the input Gaussian states of the interferometer, and the mode A is entered into the black-box undergoes the following unitary transformation  
\begin{equation}
	\hat U_A^\varphi \left( \varphi  \right) = \exp\left( {i\varphi {{\hat H}_A}} \right), \label{Eq. 13}
\end{equation}
where $\varphi$ is unknown phase shift parameter, and $\hat{H}_A$ is a generator relate to the dynamics of $A$. In order to avoid trivial dynamics,  we assume that the spectrum of  $\hat{H}_A$ is non-degenerate. After the black-box, the evolved states is write
\begin{equation}
	\rho _{AB}^\varphi  = \left( {\hat U_A^\varphi  \otimes {\mathbb{1}_B}} \right){\rho _{AB}}\left( {\hat U_A^\varphi  \otimes {\mathbb{1}_B}} \right)^{\dag}.
\end{equation}
 The quantum metrology task is the estimation of the parameter $\varphi$ with maximal possible precision. For that,  by collective processing,  one can consider an estimator $\varphi_{est}$ and improve the statistical accuracy of the estimator considered by evaluating $\Delta {\varphi_{est}^2} = \left\langle {{{\left( {{\varphi_{est}} - \varphi } \right)}^2}} \right\rangle$. Moreover, the variance of an estimator is constrained by the quantum Cramér-Rao bound \cite{cramer}
\begin{equation}
	\Delta {\varphi_{est} ^2} \ge \frac{1}{{N\,\mathcal{F}\left( {\rho _{AB}^\varphi } \right)}},\label{Eq. 9}
\end{equation}
where $N$ is the iterated number of the measurements has performed and $\mathcal{F}\left( {\rho _{AB}^\varphi } \right)$ is designated the quantum Fisher information, that can be defined as \cite{cramer, fisher}
\begin{equation}
	{\cal F}\left( {\rho _{AB}^\varphi } \right) =  - 2\mathop {\lim }\limits_{\varepsilon  \to 0} \frac{{{\partial ^2}F\left( {\rho _{AB}^\varphi ,\rho _{AB}^{\varphi  + \varepsilon }} \right)}}{{\partial {^2\varepsilon}}},
\end{equation}
where $F$ is the Uhlmann fidelity defined by \cite{uhlman}
\begin{equation}
	F\left(\rho_{1}, \rho_{2}\right)=\left\{\operatorname{Tr}\left[\left(\sqrt{\rho_{1}} \rho_{2} \sqrt{\rho_{1}}\right)^{\frac{1}{2}}\right]\right\}^{2}
\end{equation}
According to Eq. (\ref{Eq. 9}),  the limit of $N$ large ensures to reach a minimum variance of the estimator. This limit is known as the asymptotic limit \cite{hayashi2008asymptotic, kahn2009local}. Thus, the quantum Fisher information is allowed to quantify the precision in the estimation of $\varphi$ for each given choice of the black-box generator $\hat{H}_A$. Its importance consists in its inverse which attachment to the minimum variance of the optimal estimator; whenever the quantum Fisher information is large, the variance of the estimator becomes minimal. For this reason, the quantum Fisher information is adopted as the figure of merit in quantum estimation theory. In the context of estimating the parameter $\varphi$ with interest in local generator information, the minimum of the QFI is performed over all local generators. This minimization is known as Interferometry Power (IP) of the probing state $\rho_{AB}$ \cite{adesoo}
\begin{equation}
	{{\cal P}^A}\left( {{\rho _{AB}}} \right) = \frac{1}{4}\mathop {\inf }\limits_{{{\hat H}_A}} {\cal F}\left( {\rho _{AB}^{\varphi ,{{\hat H}_A}}} \right) \label{Eq. 18}
\end{equation}
Recently, it has been proved that, for general mixed states $\rho_{AB}$,  the IP is a quantum correlation measure of discord-type. It reduces to quantum entanglement in the special case of pure states \cite{pure}. The minimization of Eq. (\ref{Eq. 18}) can be solved analytically for the two-mode Gaussian states  whose covariance matrix is written in the following form;
\begin{equation}
	{\sigma _{AB}} = \left( {\begin{array}{*{20}{c}}
			\alpha &\gamma \\
			{{\gamma ^{\rm{T}}}}&\beta 
	\end{array}} \right), \label{Stand CVM}
\end{equation}
where $\alpha$ et $\beta$ are the $2 \times 2$ covariance matrices for the single modes, and $\gamma$  is a $2\times 2$  covariance matrix which contains the cross-correlations between the diverse modes. In this case,  the  minimization of Eq. (\ref{Eq. 18}) gives the expression of the GIP in terms of the covariance matrix elements \cite{12}
\begin{equation}
	\mathcal{P}_{G}^{A}\left(\sigma_{A B}\right)=\frac{X+\sqrt{X^{2}+Y Z}}{2 Y},
\end{equation}
where $X = (A + C)(1 + B + C - D) - {D^2}$,\quad $Y = (D - 1)(1 + A + B + 2C + D)$ \quad \\and  $Z = (A + D)(AB - D) + C(2A + C)(1 + B)$. The quantities $X$, $Y$, and $Z$ are expressed in terms of the symplectic invariant of the general covariance matrix of Eq. (\ref{Stand CVM}) that are given by $A=\operatorname{det} \alpha, B=\operatorname{det} \beta, C=\operatorname{det} \gamma$ and $D=\operatorname{det} \sigma_{A B}$. As mentioned in the literature, the GIP has some remarkable properties. The GIP is invariant under local unitary Gaussian operations. Also, it is monotonically non-increasing under local actions on subsystem B.  Thus, the  GIP is a suitable measure of bipartite discord-type correlations for two-mode Gaussian states \cite{adesoo}. In addition, for the general two modes, Gaussian states, the  GIP is not symmetric under exchange between mode A and mode B. The GIP reduces, for pure states, to a Gaussian entanglement monotone \cite{adessoo}.

\subsection{Gaussian entanglement of formation}
Quantum entanglement is the most significant resource in the various task of quantum information processing, so it is required to quantify this resource in many quantum systems.  In the case of quantum bipartite pure states, the Von Neumann entropy is an appropriate measure for this resource \cite{12}.  In the same direction and for more general issues, many different measurements have been introduced for two-part mixed quantum states to quantify the amount of quantum entanglement. Among these measures, one of famous interest is the entanglement of formation (EoF) \cite{bennett, wolf}. It is defined as a minimum sum of the entanglement amount of pure states
\begin{equation}
	{E_F}\left( \rho_{AB}  \right) = \min \left\{ {\sum\limits_j {{p_j}} E\left( {{\psi _j}} \right)\left| {\rho_{AB}  = \sum\limits_j {{p_j}\left| {{\psi _j}} \right\rangle \left\langle {{\psi _j}} \right|} } \right.} \right\} \label{Eq. EOF}
\end{equation}
The minimization of Eq. (\ref{Eq. EOF}) is taken over all eigenstates of the given mixed state $\rho_{AB}$, that is, in fact, the convex sum of pure states, and $ E\left( {{\psi _j}} \right)$ is the amount of entanglement of the pure state that computes using the Von Neumann entropy such as; $E\left( {{\psi _j}} \right) = S\left( {T{r_B}\left[ {\left| {{\psi _j}} \right\rangle \left\langle {{\psi _j}} \right|} \right]} \right)$.  In general, it is not easy to evaluate the minimization of Eq. (\ref{Eq. EOF}). Therefore,  the evaluation of EoF is a difficult task for the general bipartite mixed states. For this, a few analytical expressions of EoF were derived in literature, except that some particular quantum cases like; the arbitrary two-qubit states \cite{Verstraete}, and for highly symmetric states like the isotropic states \cite{B.M} and the Werner states \cite{werner}.

Gaussian family states, whose entanglement originates from squeezing modes  \cite{takei}, have played an essential role in quantum information processes of continuous-variable systems. Questions related to the EoF for the Gaussian states have been addressed. In this class of quantum states, the optimal convex decomposition involved in the definition of the entanglement of formation  (\ref{Eq. EOF}) has been solved in the special instance of two-mode symmetric mixed Gaussian states \cite{serafini} and it is given by 
\begin{equation}
	{E_F}\left( \rho  \right) = \max \left\{ {0,h\left( {{{\tilde v}_ {-} }} \right)} \right\},
\end{equation}
  where $h\left( x \right) = \frac{{{{\left( {1 + x} \right)}^2}}}{{4x}}{\log _2}\left( {\frac{{{{\left( {1 + x} \right)}^2}}}{{4x}}} \right) - \frac{{{{\left( {1 - x} \right)}^2}}}{{4x}}{\log _2}\left( {\frac{{{{\left( {1 - x} \right)}^2}}}{{4x}}} \right)$ and $\tilde{v}_{-}$ is the
  smallest symplectic eigenvalue of the partially transposed state that is given by 
  \begin{equation}
  	2{\tilde \nu _ - }^2 = \tilde \Delta ({\sigma _{AB}}) - \sqrt {\tilde \Delta {{({\sigma _{AB}})}^2} - 4{\mathop{\rm Det}\nolimits} \left( {{\sigma _{AB}}} \right)} ,
  \end{equation}
where $\tilde \Delta ({\sigma _{AB}}) = {\mathop{\rm Det}\nolimits} \alpha  + {\mathop{\rm Det}\nolimits} \beta  - 2{\mathop{\rm Det}\nolimits} \gamma$. As mentioned in \cite{vidal, serafini}, the quantification of EoF of the symmetric Gaussian states allows us to analyze the quantum correlations in an equivalence way to that one can provide by the logarithmic negativity \cite{vidal, plenio}.
\section{Results and discussion \label{sec4}}
In this section, we proposed a scheme to consider the evolution of the quantum correlations between two modes bosonic of  Gaussian-type subjected to the influence of the environment interaction. For this purpose, we initially prepare two-mode Gaussian states and make them evolve under a noisy Gaussian environment. Second, before the interaction with the environment, we submit one of the two-mode subjects to phase shift operation by the acting of the phase rotation operator. The state of the system after the interaction with the environment and phase rotation is given by $\hat{\rho}_{out}$, that we shall describe the covariance matrix $\sigma_{out}$. The proposed scheme is illustrated in Fig. (\ref{Fig. 1}).
\begin{figure}[th]

	\tikzset{every picture/.style={line width=0.75pt}} 
	
	\begin{tikzpicture}[x=0.75pt,y=0.75pt,yscale=-1,xscale=1]
		
		\draw  [color={rgb, 255:red, 38; green, 40; blue, 163 }  ,draw opacity=1 ][line width=2.25]  (3,133.5) .. controls (3,97.33) and (18.67,68) .. (38,68) .. controls (57.33,68) and (73,97.33) .. (73,133.5) .. controls (73,169.67) and (57.33,199) .. (38,199) .. controls (18.67,199) and (3,169.67) .. (3,133.5) -- cycle ;
		\draw  [color={rgb, 255:red, 38; green, 40; blue, 163 }  ,draw opacity=1 ][line width=2.25]  (133.99,193.68) -- (73,133.5) -- (134.99,70.68) ;
		\draw [color={rgb, 255:red, 38; green, 40; blue, 163 }  ,draw opacity=1 ][line width=2.25]    (132.99,193.68) -- (181.51,193.84) -- (237.5,193) ;
		\draw [color={rgb, 255:red, 38; green, 40; blue, 163 }  ,draw opacity=1 ][line width=2.25]    (134.99,71.68) -- (168.5,72) -- (159.25,71.76) ;
		\draw  [color={rgb, 255:red, 38; green, 40; blue, 163 }  ,draw opacity=1 ][line width=2.25]  (168,61.8) .. controls (168,58.04) and (171.04,55) .. (174.8,55) -- (197.7,55) .. controls (201.46,55) and (204.5,58.04) .. (204.5,61.8) -- (204.5,82.2) .. controls (204.5,85.96) and (201.46,89) .. (197.7,89) -- (174.8,89) .. controls (171.04,89) and (168,85.96) .. (168,82.2) -- cycle ;
		\draw [color={rgb, 255:red, 38; green, 40; blue, 163 }  ,draw opacity=1 ][line width=2.25]    (204.99,72.68) -- (238.5,73) -- (229.25,72.76) ;
		\draw  [color={rgb, 255:red, 38; green, 40; blue, 163 }  ,draw opacity=1 ][line width=2.25]  (514.5,130.21) .. controls (514.59,166.39) and (498.99,195.75) .. (479.66,195.8) .. controls (460.33,195.85) and (444.59,166.56) .. (444.5,130.38) .. controls (444.41,94.21) and (460.01,64.85) .. (479.34,64.8) .. controls (498.67,64.75) and (514.41,94.04) .. (514.5,130.21) -- cycle ;
		\draw  [color={rgb, 255:red, 38; green, 40; blue, 163 }  ,draw opacity=1 ][line width=2.25]  (383.37,70.35) -- (444.5,130.38) -- (382.67,193.35) ;
		\draw [color={rgb, 255:red, 38; green, 40; blue, 163 }  ,draw opacity=1 ][line width=2.25]    (383.37,70.35) -- (360.5,71) ;
		\draw [color={rgb, 255:red, 38; green, 40; blue, 163 }  ,draw opacity=1 ][line width=2.25]    (382.67,193.35) -- (358.41,193.33) -- (352.5,194) ;
		\draw  [color={rgb, 255:red, 38; green, 40; blue, 163 }  ,draw opacity=1 ][line width=2.25]  (576.99,193.68) -- (516,133.5) -- (577.99,70.68) ;
		\draw [color={rgb, 255:red, 38; green, 40; blue, 163 }  ,draw opacity=1 ][line width=2.25]    (577.99,70.68) -- (602.25,70.76) ;
		\draw [color={rgb, 255:red, 38; green, 40; blue, 163 }  ,draw opacity=1 ][line width=2.25]    (576.99,193.68) -- (601.25,193.76) ;
		\draw  [color={rgb, 255:red, 38; green, 40; blue, 163 }  ,draw opacity=1 ][line width=2.25]  (603,54) -- (618.25,54) .. controls (626.67,54) and (633.5,61.16) .. (633.5,70) .. controls (633.5,78.84) and (626.67,86) .. (618.25,86) -- (603,86) -- cycle ;
		\draw  [color={rgb, 255:red, 38; green, 40; blue, 163 }  ,draw opacity=1 ][line width=2.25]  (601,177) -- (616.25,177) .. controls (624.67,177) and (631.5,184.16) .. (631.5,193) .. controls (631.5,201.84) and (624.67,209) .. (616.25,209) -- (601,209) -- cycle ;
		\draw [color={rgb, 255:red, 38; green, 40; blue, 163 }  ,draw opacity=1 ][line width=2.25]    (633.5,70) .. controls (658,59) and (640,89) .. (668,68) ;
		\draw [color={rgb, 255:red, 38; green, 40; blue, 163 }  ,draw opacity=1 ][line width=2.25]    (631.5,193) .. controls (656,182) and (638,212) .. (666,191) ;
		\draw  [color={rgb, 255:red, 0; green, 0; blue, 0 }  ,draw opacity=1 ][fill={rgb, 255:red, 217; green, 227; blue, 240 }  ,fill opacity=1 ][line width=2.25]  (209,108) .. controls (209,91.98) and (224.67,79) .. (244,79) .. controls (263.33,79) and (279,91.98) .. (279,108) .. controls (279,124.02) and (263.33,137) .. (244,137) .. controls (224.67,137) and (209,124.02) .. (209,108) -- cycle ;
		\draw  [color={rgb, 255:red, 0; green, 0; blue, 0 }  ,draw opacity=1 ][fill={rgb, 255:red, 217; green, 227; blue, 240 }  ,fill opacity=1 ][line width=2.25]  (232,86) .. controls (232,68.88) and (254.27,55) .. (281.75,55) .. controls (309.23,55) and (331.5,68.88) .. (331.5,86) .. controls (331.5,103.12) and (309.23,117) .. (281.75,117) .. controls (254.27,117) and (232,103.12) .. (232,86) -- cycle ;
		\draw  [color={rgb, 255:red, 0; green, 0; blue, 0 }  ,draw opacity=1 ][fill={rgb, 255:red, 217; green, 227; blue, 240 }  ,fill opacity=1 ][line width=2.25]  (280,76) .. controls (280,59.98) and (295.67,47) .. (315,47) .. controls (334.33,47) and (350,59.98) .. (350,76) .. controls (350,92.02) and (334.33,105) .. (315,105) .. controls (295.67,105) and (280,92.02) .. (280,76) -- cycle ;
		\draw  [color={rgb, 255:red, 0; green, 0; blue, 0 }  ,draw opacity=1 ][fill={rgb, 255:red, 217; green, 227; blue, 240 }  ,fill opacity=1 ][line width=2.25]  (303,96) .. controls (303,77.22) and (318.67,62) .. (338,62) .. controls (357.33,62) and (373,77.22) .. (373,96) .. controls (373,114.78) and (357.33,130) .. (338,130) .. controls (318.67,130) and (303,114.78) .. (303,96) -- cycle ;
		\draw  [color={rgb, 255:red, 0; green, 0; blue, 0 }  ,draw opacity=1 ][fill={rgb, 255:red, 217; green, 227; blue, 240 }  ,fill opacity=1 ][line width=2.25]  (316,120) .. controls (316,103.98) and (331.67,91) .. (351,91) .. controls (370.33,91) and (386,103.98) .. (386,120) .. controls (386,136.02) and (370.33,149) .. (351,149) .. controls (331.67,149) and (316,136.02) .. (316,120) -- cycle ;
		\draw  [color={rgb, 255:red, 0; green, 0; blue, 0 }  ,draw opacity=1 ][fill={rgb, 255:red, 217; green, 227; blue, 240 }  ,fill opacity=1 ][line width=2.25]  (312,145) .. controls (312,128.98) and (327.67,116) .. (347,116) .. controls (366.33,116) and (382,128.98) .. (382,145) .. controls (382,161.02) and (366.33,174) .. (347,174) .. controls (327.67,174) and (312,161.02) .. (312,145) -- cycle ;
		\draw  [color={rgb, 255:red, 0; green, 0; blue, 0 }  ,draw opacity=1 ][fill={rgb, 255:red, 217; green, 227; blue, 240 }  ,fill opacity=1 ][line width=2.25]  (309,168) .. controls (309,151.98) and (324.67,139) .. (344,139) .. controls (363.33,139) and (379,151.98) .. (379,168) .. controls (379,184.02) and (363.33,197) .. (344,197) .. controls (324.67,197) and (309,184.02) .. (309,168) -- cycle ;
		\draw  [color={rgb, 255:red, 0; green, 0; blue, 0 }  ,draw opacity=1 ][fill={rgb, 255:red, 217; green, 227; blue, 240 }  ,fill opacity=1 ][line width=2.25]  (284,185) .. controls (284,168.98) and (299.67,156) .. (319,156) .. controls (338.33,156) and (354,168.98) .. (354,185) .. controls (354,201.02) and (338.33,214) .. (319,214) .. controls (299.67,214) and (284,201.02) .. (284,185) -- cycle ;
		\draw  [color={rgb, 255:red, 0; green, 0; blue, 0 }  ,draw opacity=1 ][fill={rgb, 255:red, 217; green, 227; blue, 240 }  ,fill opacity=1 ][line width=2.25]  (209,152) .. controls (209,135.98) and (224.67,123) .. (244,123) .. controls (263.33,123) and (279,135.98) .. (279,152) .. controls (279,168.02) and (263.33,181) .. (244,181) .. controls (224.67,181) and (209,168.02) .. (209,152) -- cycle ;
		\draw  [color={rgb, 255:red, 0; green, 0; blue, 0 }  ,draw opacity=1 ][fill={rgb, 255:red, 217; green, 227; blue, 240 }  ,fill opacity=1 ][line width=2.25]  (232,184) .. controls (232,167.98) and (247.67,155) .. (267,155) .. controls (286.33,155) and (302,167.98) .. (302,184) .. controls (302,200.02) and (286.33,213) .. (267,213) .. controls (247.67,213) and (232,200.02) .. (232,184) -- cycle ;
		\draw  [color={rgb, 255:red, 0; green, 0; blue, 0 }  ,draw opacity=1 ][fill={rgb, 255:red, 217; green, 227; blue, 240 }  ,fill opacity=1 ][line width=2.25]  (223.5,129.5) .. controls (223.5,90.01) and (256.63,58) .. (297.5,58) .. controls (338.37,58) and (371.5,90.01) .. (371.5,129.5) .. controls (371.5,168.99) and (338.37,201) .. (297.5,201) .. controls (256.63,201) and (223.5,168.99) .. (223.5,129.5) -- cycle ;
		\draw  [color={rgb, 255:red, 217; green, 227; blue, 240 }  ,draw opacity=1 ][line width=3] [line join = round][line cap = round] (278.5,62) .. controls (264.28,64.84) and (247.22,74.28) .. (239.5,82) .. controls (235.96,85.54) and (234.04,90.46) .. (230.5,94) ;
		\draw  [color={rgb, 255:red, 217; green, 227; blue, 240 }  ,draw opacity=1 ][line width=3] [line join = round][line cap = round] (260.5,72) .. controls (246.72,72) and (227.63,95.61) .. (226.5,108) .. controls (225.93,114.31) and (226.82,120.67) .. (226.5,127) .. controls (226.18,133.47) and (223.5,139.69) .. (223.5,146) ;
		\draw  [color={rgb, 255:red, 217; green, 227; blue, 240 }  ,draw opacity=1 ][line width=3] [line join = round][line cap = round] (244.5,79) .. controls (231.54,79) and (228.24,95.52) .. (224.5,103) ;
		\draw  [color={rgb, 255:red, 217; green, 227; blue, 240 }  ,draw opacity=1 ][line width=3] [line join = round][line cap = round] (247.5,82) .. controls (234.1,82) and (228.5,102.26) .. (228.5,113) ;
		\draw  [color={rgb, 255:red, 217; green, 227; blue, 240 }  ,draw opacity=1 ][line width=3] [line join = round][line cap = round] (287.5,61) .. controls (276.15,61) and (256.02,65.48) .. (248.5,73) .. controls (242.68,78.82) and (240.39,88.81) .. (236.5,94) .. controls (226.28,107.63) and (214.27,120.77) .. (227.5,134) ;
		\draw  [color={rgb, 255:red, 217; green, 227; blue, 240 }  ,draw opacity=1 ][line width=3] [line join = round][line cap = round] (227.5,112) .. controls (227.5,114.4) and (224.5,119.31) .. (224.5,124) ;
		\draw  [color={rgb, 255:red, 217; green, 227; blue, 240 }  ,draw opacity=1 ][line width=3] [line join = round][line cap = round] (231.5,127) .. controls (219.69,127) and (218.46,139.95) .. (226.5,150) ;
		\draw  [color={rgb, 255:red, 217; green, 227; blue, 240 }  ,draw opacity=1 ][line width=3] [line join = round][line cap = round] (231.5,134) .. controls (231.5,135.41) and (229.23,135.79) .. (228.5,137) .. controls (225.41,142.15) and (227.1,149) .. (227.5,155) .. controls (227.88,160.76) and (235.62,165.24) .. (237.5,169) .. controls (239.34,172.69) and (240.48,178.48) .. (243.5,183) ;
		\draw  [color={rgb, 255:red, 217; green, 227; blue, 240 }  ,draw opacity=1 ][line width=3] [line join = round][line cap = round] (228.5,147) .. controls (227.55,147.95) and (221.64,151.51) .. (221.5,152) .. controls (220.49,155.53) and (221.17,159.35) .. (221.5,163) .. controls (221.89,167.26) and (241.02,161.66) .. (234.5,157) .. controls (232.6,155.64) and (229.83,157) .. (227.5,157) ;
		\draw  [color={rgb, 255:red, 217; green, 227; blue, 240 }  ,draw opacity=1 ][line width=3] [line join = round][line cap = round] (225.5,160) .. controls (225.5,162.69) and (224,167) .. (226.5,168) .. controls (230.49,169.6) and (231.78,168.28) .. (235.5,172) .. controls (237.5,174) and (238.67,178) .. (241.5,178) ;
		\draw  [color={rgb, 255:red, 217; green, 227; blue, 240 }  ,draw opacity=1 ][line width=3] [line join = round][line cap = round] (235.5,148) .. controls (216.2,148) and (237.53,171.03) .. (241.5,175) ;
		\draw  [color={rgb, 255:red, 217; green, 227; blue, 240 }  ,draw opacity=1 ][line width=3] [line join = round][line cap = round] (242.5,173) .. controls (242.5,186.52) and (262.99,186.2) .. (272.5,190) .. controls (274.83,190.93) and (281.67,197.66) .. (280.5,200) .. controls (273.91,213.19) and (262.9,193.3) .. (265.5,192) .. controls (267.43,191.04) and (276.15,189.69) .. (273.5,195) .. controls (271.93,198.14) and (259.22,197.17) .. (258.5,195) .. controls (258.18,194.05) and (257.51,192.16) .. (258.5,192) .. controls (264.76,191.01) and (271.2,191.34) .. (277.5,192) .. controls (278.16,192.07) and (278.16,193.88) .. (277.5,194) .. controls (273.89,194.66) and (270.15,194.33) .. (266.5,194) .. controls (265.5,193.91) and (266.14,190.16) .. (264.5,190) .. controls (261.51,189.7) and (258.5,190) .. (255.5,190) .. controls (254.45,190) and (257.45,189.12) .. (258.5,189) .. controls (261.15,188.71) and (263.85,188.71) .. (266.5,189) .. controls (268.64,189.24) and (268.29,199.74) .. (268.5,200) .. controls (271.42,203.51) and (273.31,199) .. (275.5,199) ;
		\draw  [color={rgb, 255:red, 217; green, 227; blue, 240 }  ,draw opacity=1 ][line width=3] [line join = round][line cap = round] (291.5,198) .. controls (283.8,198) and (276.16,200.77) .. (268.5,200) .. controls (263.11,199.46) and (264.01,194.78) .. (260.5,194) .. controls (258.22,193.49) and (255.82,194.29) .. (253.5,194) .. controls (251.38,193.74) and (250.6,190.83) .. (249.5,189) .. controls (248.36,187.1) and (249.98,186.24) .. (251.5,187) .. controls (254.27,188.39) and (252.42,192.56) .. (255.5,193) .. controls (257.22,193.25) and (277.18,192.52) .. (279.5,196) .. controls (281.43,198.9) and (276.12,202.62) .. (274.5,201) .. controls (274.34,200.84) and (273.62,198.06) .. (273.5,198) .. controls (272.46,197.48) and (270.96,198.84) .. (270.5,197) .. controls (270.21,195.83) and (272.5,192.8) .. (272.5,194) .. controls (272.5,197.41) and (262.69,196) .. (266.5,196) ;
		\draw  [color={rgb, 255:red, 217; green, 227; blue, 240 }  ,draw opacity=1 ][line width=3] [line join = round][line cap = round] (372.5,128) .. controls (359.83,140.67) and (365.84,152.84) .. (362.5,169) .. controls (361.99,171.48) and (354.52,177.31) .. (352.5,180) .. controls (343.45,192.07) and (316.58,201) .. (298.5,201) ;
		\draw  [color={rgb, 255:red, 217; green, 227; blue, 240 }  ,draw opacity=1 ][line width=3] [line join = round][line cap = round] (371.5,156) .. controls (356.36,156) and (350.15,171.35) .. (342.5,179) .. controls (337.79,183.71) and (334.74,186.14) .. (331.5,191) .. controls (331.09,191.62) and (329.77,192.85) .. (330.5,193) .. controls (338.45,194.59) and (351.17,180.17) .. (357.5,177) .. controls (357.54,176.98) and (369.89,170.13) .. (366.5,169) .. controls (361.95,167.48) and (347.43,177.13) .. (345.5,181) .. controls (344.75,182.49) and (348.42,179.27) .. (349.5,178) .. controls (352.97,173.9) and (355.1,168.81) .. (357.5,164) .. controls (357.87,163.26) and (362.34,160.84) .. (361.5,160) .. controls (361.26,159.76) and (360.8,159.85) .. (360.5,160) .. controls (355.79,162.35) and (350.7,170.27) .. (346.5,174) .. controls (344.03,176.19) and (341.98,178.83) .. (339.5,181) .. controls (338.94,181.49) and (336.97,182.53) .. (337.5,182) .. controls (341.54,177.96) and (352.04,171.66) .. (355.5,169) .. controls (356.45,168.27) and (359.48,166.3) .. (358.5,167) .. controls (352.03,171.62) and (346.03,176.43) .. (339.5,181) .. controls (338.34,181.81) and (335.32,183.22) .. (336.5,184) .. controls (338.93,185.62) and (359.5,176.59) .. (359.5,174) .. controls (359.5,171.57) and (354.67,174.91) .. (352.5,176) .. controls (350.33,177.09) and (347.72,177.01) .. (345.5,178) .. controls (343.55,178.87) and (342.31,180.87) .. (340.5,182) .. controls (339.06,182.9) and (333.8,183) .. (335.5,183) .. controls (336.66,183) and (344.5,183.5) .. (344.5,181) .. controls (344.5,179.33) and (341.16,180.82) .. (339.5,181) .. controls (334.73,181.53) and (331.84,184.83) .. (329.5,186) .. controls (328.43,186.54) and (325.31,187.8) .. (326.5,188) .. controls (328.82,188.39) and (331.2,187.51) .. (333.5,187) .. controls (334.53,186.77) and (337.25,185.25) .. (336.5,186) .. controls (336.12,186.38) and (328.76,189.63) .. (329.5,190) .. controls (330.73,190.61) and (334.47,189.97) .. (333.5,189) .. controls (331.49,186.99) and (326.07,192.57) .. (325.5,192) .. controls (325.45,191.95) and (329.62,187.97) .. (333.5,187) .. controls (334.22,186.82) and (336.17,186.33) .. (335.5,186) .. controls (334.9,185.7) and (334.1,185.72) .. (333.5,186) .. controls (329.04,188.08) and (324.9,190.8) .. (320.5,193) .. controls (318.89,193.8) and (316.3,193.39) .. (315.5,195) .. controls (314.9,196.19) and (318.18,195.18) .. (319.5,195) .. controls (324.95,194.27) and (330.28,192.74) .. (335.5,191) .. controls (337.63,190.29) and (325.28,194) .. (327.5,194) .. controls (332.67,194) and (337.72,191.99) .. (342.5,190) .. controls (342.89,189.84) and (352.7,185.08) .. (352.5,185) .. controls (346.73,182.69) and (345.31,185.6) .. (342.5,187) .. controls (341.27,187.61) and (337.14,187.77) .. (338.5,188) .. controls (341.48,188.5) and (349.63,189.13) .. (347.5,187) .. controls (347.03,186.53) and (344.83,187) .. (345.5,187) .. controls (348.17,187) and (350.83,187) .. (353.5,187) .. controls (361.54,187) and (337.71,186.11) .. (343.5,189) .. controls (346.25,190.37) and (349.75,188.37) .. (352.5,187) .. controls (353.73,186.39) and (357.73,186.61) .. (356.5,186) .. controls (354.98,185.24) and (349.85,186.59) .. (351.5,187) .. controls (353.12,187.4) and (354.83,187) .. (356.5,187) .. controls (357.83,187) and (353.83,187) .. (352.5,187) ;
		\draw  [color={rgb, 255:red, 217; green, 227; blue, 240 }  ,draw opacity=1 ][line width=3] [line join = round][line cap = round] (360.5,178) .. controls (360.5,180.67) and (361.02,183.39) .. (360.5,186) .. controls (359.82,189.38) and (357.29,187.28) .. (355.5,188) .. controls (354.62,188.35) and (354.02,190.78) .. (353.5,190) .. controls (352.58,188.61) and (353.04,186.6) .. (353.5,185) .. controls (353.76,184.09) and (354.83,183.67) .. (355.5,183) ;
		\draw  [color={rgb, 255:red, 217; green, 227; blue, 240 }  ,draw opacity=1 ][line width=3] [line join = round][line cap = round] (359.5,157) .. controls (359.5,168.04) and (357.84,176.66) .. (350.5,184) .. controls (349.5,185) and (352.65,182.13) .. (353.5,181) .. controls (356.71,176.72) and (359.85,172.41) .. (362.5,168) .. controls (366.95,160.58) and (376.8,138.08) .. (373.5,129) .. controls (372.63,126.61) and (367.91,129.89) .. (366.5,132) .. controls (363.62,136.33) and (363.41,155.91) .. (365.5,158) .. controls (367.35,159.85) and (372.42,156.22) .. (372.5,154) .. controls (372.85,144.29) and (368.63,131.87) .. (375.5,125) .. controls (377.95,122.55) and (371.18,126.55) .. (370.5,127) .. controls (367.28,129.15) and (364.3,130.18) .. (365.5,135) .. controls (365.76,136.02) and (374.57,130.7) .. (375.5,130) .. controls (376.1,129.55) and (378.25,129) .. (377.5,129) .. controls (375.26,129) and (373.62,131.29) .. (371.5,132) .. controls (370.5,132.33) and (369.25,132.25) .. (368.5,133) ;
		\draw  [color={rgb, 255:red, 217; green, 227; blue, 240 }  ,draw opacity=1 ][line width=3] [line join = round][line cap = round] (373.5,139) .. controls (370.81,139) and (369.02,151.19) .. (368.5,153) .. controls (368.11,154.36) and (365.58,157.41) .. (365.5,156) .. controls (365.09,149.1) and (364.19,141.61) .. (366.5,137) .. controls (368.3,133.41) and (366.63,145.08) .. (367.5,149) .. controls (367.82,150.46) and (368.45,146.05) .. (369.5,145) .. controls (370.87,143.63) and (372.5,141.94) .. (372.5,140) .. controls (372.5,138.3) and (371.67,143.31) .. (371.5,145) .. controls (370.82,151.82) and (370.58,153.5) .. (369.5,160) .. controls (368.31,167.15) and (358.78,173.15) .. (356.5,180) .. controls (355.64,182.58) and (354.77,187.18) .. (352.5,189) .. controls (349.47,191.43) and (340.59,196.09) .. (337.5,193) ;
		\draw  [color={rgb, 255:red, 217; green, 227; blue, 240 }  ,draw opacity=1 ][line width=3] [line join = round][line cap = round] (372.5,109) .. controls (372.5,116.33) and (372.91,123.68) .. (372.5,131) .. controls (372.39,133.02) and (371.99,126.97) .. (371.5,125) .. controls (370.44,120.75) and (369.94,116.36) .. (369.5,112) .. controls (368.68,103.79) and (361.86,97.54) .. (357.5,91) .. controls (357.09,90.38) and (358.24,92.3) .. (358.5,93) .. controls (359.29,95.09) and (360.5,97) .. (361.5,99) ;
		\draw  [color={rgb, 255:red, 217; green, 227; blue, 240 }  ,draw opacity=1 ][line width=3] [line join = round][line cap = round] (367.5,103) .. controls (367.5,101.67) and (367.67,100.32) .. (367.5,99) .. controls (367.33,97.64) and (367.11,96.23) .. (366.5,95) .. controls (366.08,94.16) and (364.8,92.11) .. (364.5,93) .. controls (364.08,94.26) and (365.1,95.81) .. (364.5,97) .. controls (363.75,98.49) and (361.25,98.51) .. (360.5,100) .. controls (360.37,100.26) and (359.19,101.69) .. (358.5,101) .. controls (358.26,100.76) and (358.35,100.3) .. (358.5,100) .. controls (360.87,95.27) and (361.32,91.18) .. (364.5,88) .. controls (364.6,87.9) and (367.5,103.45) .. (367.5,105) ;
		\draw  [color={rgb, 255:red, 217; green, 227; blue, 240 }  ,draw opacity=1 ][line width=3] [line join = round][line cap = round] (352.5,84) .. controls (355.97,87.47) and (353.9,80.69) .. (351.5,80) .. controls (349.9,79.54) and (348.15,80.24) .. (346.5,80) .. controls (345.31,79.83) and (343.38,77.45) .. (344.5,77) .. controls (349.55,74.98) and (358.97,79.53) .. (362.5,82) .. controls (364.23,83.21) and (366.47,87.26) .. (364.5,88) .. controls (361.67,89.06) and (358.48,87.5) .. (355.5,87) .. controls (353.5,86.67) and (348.07,86.57) .. (349.5,88) .. controls (351.3,89.8) and (351.56,83.36) .. (352.5,81) .. controls (352.75,80.38) and (354.25,80.38) .. (354.5,81) .. controls (356.26,85.4) and (357.5,91.95) .. (357.5,97) ;
		\draw  [color={rgb, 255:red, 217; green, 227; blue, 240 }  ,draw opacity=1 ][line width=3] [line join = round][line cap = round] (371.5,131) .. controls (371.5,146.69) and (363.13,157.75) .. (356.5,171) ;
		\draw  [color={rgb, 255:red, 217; green, 227; blue, 240 }  ,draw opacity=1 ][line width=3] [line join = round][line cap = round] (283.5,197) .. controls (277.71,198.16) and (274.15,199.21) .. (267.5,198) .. controls (265.86,197.7) and (265.99,194.75) .. (264.5,194) .. controls (259.02,191.26) and (256.84,190.05) .. (249.5,189) .. controls (248.76,188.89) and (250.76,188.08) .. (251.5,188) .. controls (254.15,187.71) and (256.85,187.73) .. (259.5,188) .. controls (264.5,188.5) and (274.3,192.8) .. (277.5,196) ;
		\draw  [color={rgb, 255:red, 217; green, 227; blue, 240 }  ,draw opacity=1 ][line width=3] [line join = round][line cap = round] (352.5,81) .. controls (338.23,75.29) and (327.08,65) .. (311.5,65) .. controls (309.39,65) and (315.5,63.67) .. (317.5,63) .. controls (320.38,62.04) and (330.21,61.36) .. (331.5,62) .. controls (334.37,63.43) and (334.67,68.5) .. (335.5,71) .. controls (336.2,73.11) and (341.54,73.82) .. (343.5,75) .. controls (343.74,75.15) and (344.73,76.77) .. (345.5,76) .. controls (347.17,74.33) and (342.17,72.67) .. (340.5,71) .. controls (339.32,69.82) and (339.15,67.21) .. (337.5,67) .. controls (334.85,66.67) and (332.1,67.58) .. (329.5,67) .. controls (328.77,66.84) and (329.21,65.24) .. (328.5,65) .. controls (326.58,64.36) and (320.47,66) .. (322.5,66) .. controls (328.36,66) and (332.32,63.04) .. (336.5,62) .. controls (337.47,61.76) and (339.32,61.02) .. (339.5,62) .. controls (340.48,67.24) and (339.8,72.67) .. (339.5,78) .. controls (339.26,82.31) and (335.5,85.96) .. (335.5,90) ;
		\draw  [color={rgb, 255:red, 217; green, 227; blue, 240 }  ,draw opacity=1 ][line width=3] [line join = round][line cap = round] (326.5,60) .. controls (320.98,60) and (316.64,62.71) .. (311.5,64) .. controls (308.76,64.69) and (306.05,63.73) .. (303.5,65) .. controls (302.56,65.47) and (299.5,66.33) .. (300.5,66) .. controls (304.11,64.8) and (307.94,64.33) .. (311.5,63) .. controls (313.59,62.21) and (319.74,60) .. (317.5,60) .. controls (311.17,60) and (304.83,59.68) .. (298.5,60) .. controls (297.13,60.07) and (293.13,61) .. (294.5,61) .. controls (299.43,61) and (300.4,61.27) .. (305.5,60) .. controls (306.22,59.82) and (308.21,59.24) .. (307.5,59) .. controls (305.8,58.43) and (297.67,59.42) .. (296.5,60) .. controls (295.61,60.45) and (307.58,59) .. (305.5,59) .. controls (302.17,59) and (298.82,59.33) .. (295.5,59) .. controls (294.29,58.88) and (297.83,57.08) .. (297.5,57) .. controls (295.21,56.43) and (291.07,55.71) .. (290.5,58) .. controls (289.56,61.74) and (291.03,64.53) .. (293.5,67) ;
		\draw  [color={rgb, 255:red, 217; green, 227; blue, 240 }  ,draw opacity=1 ][line width=3] [line join = round][line cap = round] (323.5,57) .. controls (306.72,57) and (306.99,58.32) .. (294.5,63) .. controls (291.33,64.19) and (289.06,60.8) .. (287.5,66) .. controls (286.83,68.23) and (287.5,70.67) .. (287.5,73) ;
		\draw  [color={rgb, 255:red, 217; green, 227; blue, 240 }  ,draw opacity=1 ][line width=3] [line join = round][line cap = round] (310.5,55) .. controls (302.56,55) and (298.41,58.03) .. (292.5,60) .. controls (277.6,64.97) and (279.42,75.45) .. (276.5,93) ;
		\draw  [color={rgb, 255:red, 217; green, 227; blue, 240 }  ,draw opacity=1 ][line width=3] [line join = round][line cap = round] (293.5,60) .. controls (293.5,60) and (293.5,60) .. (293.5,60) ;
		\draw  [color={rgb, 255:red, 217; green, 227; blue, 240 }  ,draw opacity=1 ][line width=3] [line join = round][line cap = round] (293.5,59) .. controls (282.12,59) and (258.38,66.24) .. (253.5,76) ;
		\draw  [color={rgb, 255:red, 217; green, 227; blue, 240 }  ,draw opacity=1 ][line width=3] [line join = round][line cap = round] (291.5,59) .. controls (287.52,59) and (282.79,61.34) .. (279.5,62) .. controls (270.58,63.78) and (261.5,59.58) .. (261.5,73) ;
		\draw  [color={rgb, 255:red, 217; green, 227; blue, 240 }  ,draw opacity=1 ][line width=3] [line join = round][line cap = round] (284.5,59) .. controls (265.68,59) and (264.5,58.3) .. (264.5,76) ;
		\draw  [color={rgb, 255:red, 217; green, 227; blue, 240 }  ,draw opacity=1 ][line width=3] [line join = round][line cap = round] (376.5,152) .. controls (376.5,152) and (376.5,152) .. (376.5,152) ;
		\draw  [color={rgb, 255:red, 217; green, 227; blue, 240 }  ,draw opacity=1 ][line width=3] [line join = round][line cap = round] (376.5,151) .. controls (375.06,151) and (373.58,153.38) .. (372.5,155) ;
		\draw  [color={rgb, 255:red, 217; green, 227; blue, 240 }  ,draw opacity=1 ][line width=3] [line join = round][line cap = round] (364.5,95) .. controls (362.14,95) and (358.15,96.27) .. (357.5,94) .. controls (356,88.74) and (358.82,83.76) .. (360.5,86) .. controls (361.7,87.6) and (360.22,90.02) .. (360.5,92) .. controls (361.38,98.13) and (367.5,106.87) .. (367.5,93) ;
		\draw  [color={rgb, 255:red, 217; green, 227; blue, 240 }  ,draw opacity=1 ][line width=3] [line join = round][line cap = round] (231.5,85) .. controls (231.5,85) and (231.5,85) .. (231.5,85) ;
		\draw  [color={rgb, 255:red, 217; green, 227; blue, 240 }  ,draw opacity=1 ][line width=3] [line join = round][line cap = round] (287.5,202) .. controls (287.5,202) and (287.5,202) .. (287.5,202) ;
		\draw  [color={rgb, 255:red, 217; green, 227; blue, 240 }  ,draw opacity=1 ][line width=3] [line join = round][line cap = round] (288.5,198) .. controls (288.37,198) and (277.89,200.7) .. (281.5,201) .. controls (285.83,201.36) and (290.28,201.05) .. (294.5,200) .. controls (295.52,199.74) and (296.5,199.33) .. (297.5,199) .. controls (298.13,198.79) and (300.17,199) .. (299.5,199) .. controls (294.17,199) and (278.17,199) .. (283.5,199) .. controls (290.17,199) and (296.83,199) .. (303.5,199) .. controls (306.19,199) and (295.5,200.69) .. (295.5,198) ;
		\draw  [color={rgb, 255:red, 217; green, 227; blue, 240 }  ,draw opacity=1 ][line width=3] [line join = round][line cap = round] (296.5,203) .. controls (296.5,203) and (296.5,203) .. (296.5,203) ;
		\draw  [color={rgb, 255:red, 217; green, 227; blue, 240 }  ,draw opacity=1 ][line width=3] [line join = round][line cap = round] (234.5,175) .. controls (233.01,175) and (230.5,174.49) .. (230.5,173) ;

		\draw (39,133) node  [font=\LARGE,color={rgb, 255:red, 138; green, 5; blue, 22 }  ,opacity=1 ,rotate=-359.87]  {$\hat \rho _{inp}$};
		\draw (296,124) node  [font=\Large,color={rgb, 255:red, 156; green, 5; blue, 23 }  ,opacity=1 ,rotate=-331.94]  {$ \begin{array}{l}
				{\textstyle \mathbf{\ \ \ Gaussian\ }}\\
				\mathbf{{\textstyle Environment}}
			\end{array}$};
		\draw (186.25,71) node  [font=\LARGE,color={rgb, 255:red, 148; green, 17; blue, 7 }  ,opacity=1 ,rotate=-358.59]  {$\varphi$};
		\draw (482,133) node  [font=\LARGE,color={rgb, 255:red, 139; green, 9; blue, 25 }  ,opacity=1 ,rotate=-359.87]  {$\hat \rho _{out}$};

	\end{tikzpicture}
\captionof{figure}{A schematic illustrates the protocol proposed.  Via the GIP and the EoF, the proposed protocol aims to study the dynamics of quantum correlation of a two-mode Gaussian state under the noisy Gaussian environment. We prepare first the probe state $\hat \rho_{inp}$, using the general two-mode Gaussian state as described by the Eq. (\ref{EQ. 24}). Next, before the interaction with the Gaussian environment, one of the input modes is subjected to an unknown Gaussian unitary transformation modeled by $U_A^{\varphi}$ given in Eq. (\ref{Eq. 13}). After this unitary evolution, the two modes are evolving under a Gaussian thermal environment according to the master equation (\ref{Eq. 9}), and the final output state is given by the covariance matrix (\ref{EQ. 27}).}\label{Fig. 1}
\end{figure}
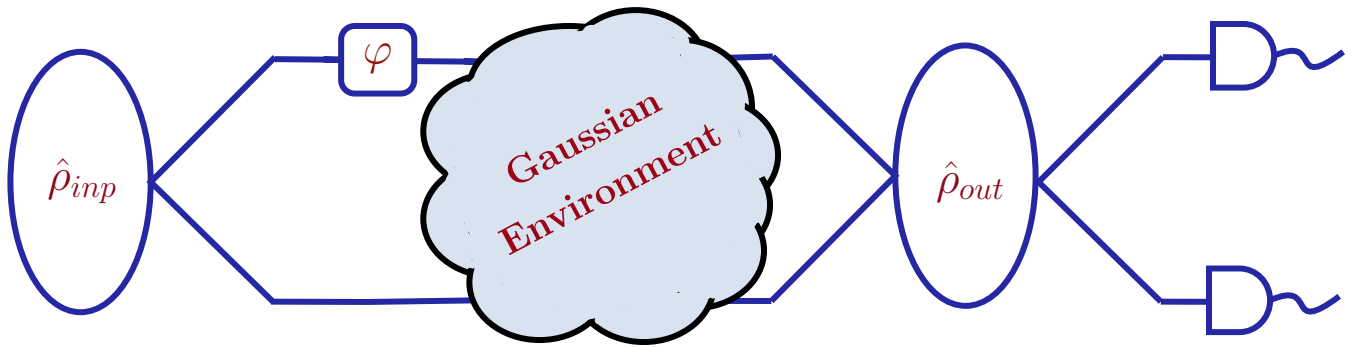	

In this work, we choose a two-mode squeezed rotated displaced thermal state as an input state, which is the most general two-mode state, with the initial squeezing parameter $r$, the rotation angle $\theta$, and the initial displacement $\boldsymbol{\alpha}$ that is a parameter characterizing the coherent light. The matrix density of this input state is given by
\begin{equation}
{{\hat \rho }_{inp}} = {{\hat D}_2}({\boldsymbol{\alpha}}){{\hat R}_2}\left( \theta  \right){{\hat S}_2}(r )\left( {{{\hat \rho }_{th}} \otimes {{\hat \rho }_{th}}} \right){{\hat S}_2}{(r )^\dag }{{\hat R}_2}{\left( \theta  \right)^\dag }{{\hat D}_2}{({\boldsymbol{\alpha}})^\dag },\label{EQ. 24}
\end{equation}
where ${{\hat S}_2}(r) = \exp \left( {r\left( {\hat a_1^\dag \hat a_2^\dag  - {{\hat a}_1}{{\hat a}_2}} \right)} \right)$ is the two mode squeezing operator with the squeezed parameter $r$, and ${{\hat R}_2}\left( \theta  \right) = \exp \left( {\theta \left( {a_1^\dag {a_2} - {a_1}a_2^\dag } \right)} \right)$ is the two-mode rotation operator in the phase space, with the rotation angle $\theta$. It also corresponds to the Beamer splitter that is characterized by the transmissivity $\tau=\cos^{2} \theta$. The two-mode displacement operator ${{\hat D}_2}$ is given such as ${{\hat D}_2}\left( \boldsymbol{\alpha}  \right) = \exp \left( {{\alpha _1}\hat a_1^\dag  - \alpha _1^*{{\hat a}_1} + {\alpha _2}\hat a_2^\dag  - \alpha _2^*{{\hat a}_2}} \right)$, with the parameter $\boldsymbol{\alpha}$. In Eq. (\ref{EQ. 24}), $\hat \rho_{th}$ denote the thermal state, which is
given by
\begin{equation}
	{{\hat \rho }_{th}} = \sum\limits_n {\frac{{{{\bar n}^n}}}{{{{(\bar n + 1)}^{n + 1}}}}} |n\rangle \langle n|,
\end{equation}
where $\bar{n}_i=\operatorname{Tr}\left(\hat{\rho}_{t h} \hat{a_i}^{\dagger} \hat{a_i}\right)$ is  the mean number of photons in $i^{th}$ bosonic mode, which is expressed in terms of the temperature effect as $\bar{n}_i=\left(e^{\beta}-1\right)^{-1}$. At zero temperature limits, the two-mode thermal state becomes the two-mode pure vacuum state.\\
From Eq. (\ref{EQ. 24}), one can easily check that the corresponding Wigner function or characteristic function  is Gaussian with the  covariance matrix
\begin{equation}
	{\sigma _{{\rm{inp }}}} = {{\hat R}_2}(\theta ){{\hat S}_2}(r )\left( {2{{\bar n}_1} + 1} \right)\mathbb{1}_{2\times 2} \oplus \left( {2{{\bar n}_2} + 1} \right)\mathbb{1}_{2\times 2}{{\hat S}_2}{(r )^\dag }{{\hat R}_2}{(\theta )^\dag }
\end{equation}
where ${{\hat R}_2}(\theta )$ and ${{\hat S}_2}(r )$ are the symplectic transformations related, respectively, to the squeezing and rotation operators. The two-mode bosonic system evolves under a noisy Gaussian non-unitary channel, and it is described by the master equation  (\ref{equ1}). The corresponding Winger representation is given in terms of the covariance matrix 
\begin{equation}
	{\sigma _{{\rm{out }}}}(t) = {e^{ - \gamma t}}\left( {U_A^\varphi  \oplus {\mathbb{1}_B}} \right){\sigma _{{\rm{inp}}}}{\left( {U_A^\varphi  \oplus {\mathbb{1}_B}} \right)^\dag } + \left( {1 - {e^{ - \gamma t}}} \right)\sigma_{\infty}. \label{EQ. 27}
\end{equation}
Since the considered environment is  Gaussian type and  when the squeezing parameter of baths vanishes,  the diffusion matrix $\sigma_{\infty}$ is diagonal and takes the following form
\begin{equation}
	{\sigma _\infty } = \left( {2{{\bar n}_1+1}} \right){\mathbb{1}_{2 \times 2}} \oplus \left( {2{{\bar n}_2} + 1} \right){\mathbb{1}_{2 \times 2}}.
\end{equation}

Now, we arrive at the study of the degree of entanglement and the quantum correlation between the two output modes. These modes are described by the covariance matrix given in Eq. (\ref{EQ. 27}). In the accomplishments of this study,  we employ two divers quantifiers:  Gaussian interferometer power (GIP) and the Gaussian entanglement of formation (EoF). The results obtained numerically are presented in the different figures below. The different figures are plotted in terms of the various parameters of the initial state.  In addition, to clarify the dynamic of quantum correlations during the interaction with the environment, we plotted the temporal evolution behavior of Gaussian interferometer power and the Gaussian entanglement of formation.

\begin{figure}[H]
	\centering
	\begin{subfigure}{.49\textwidth}
		\centering
		\includegraphics[width=8cm]{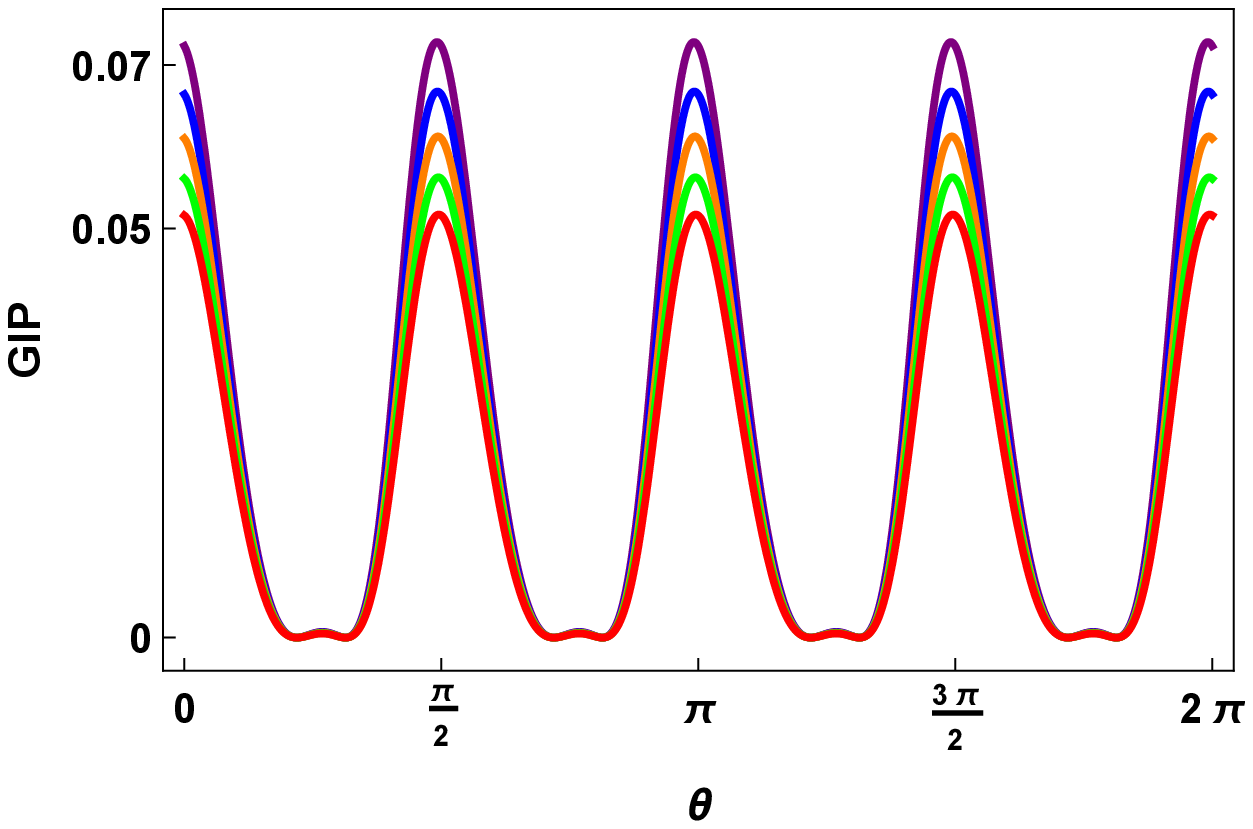}
		\caption{}\label{Fig. 2a}
	\end{subfigure}
	\begin{subfigure}{.49\textwidth}
		\centering
		\includegraphics[width=8cm]{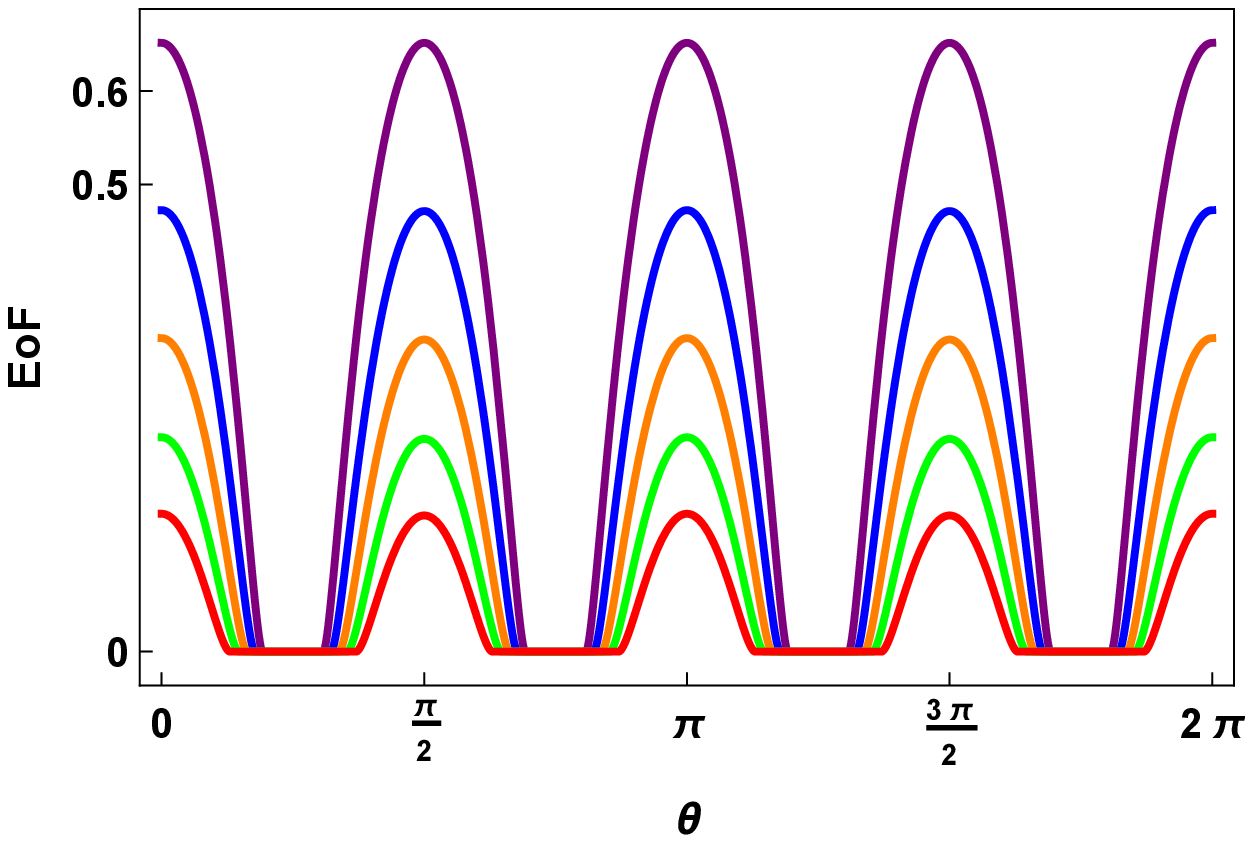}
		\caption{}\label{Fig. 2b}
	\end{subfigure}
	\renewcommand\thefigure{\arabic{figure}}
	\addtocounter{figure}{-1}
	\captionsetup{justification=raggedright, singlelinecheck=false, labelfont=sc} \captionof{figure}{The plot of Gaussian interferometric power and Gaussian entanglement of formation, for a two-mode output state, as a function of the rotation angle $\theta$. Fig. (\ref{Fig. 2a}) represents the behavior of GIP as a function of $\theta$ for the different values of time $t$, with $\bar{n}_1=\bar{n}_2= 0,5$ is the mean number of a photon of the Gaussian environment, and the initial squeezing parameter taking the value 0.3. Fig. (\ref{Fig. 2b}) represents the dynamic of EoF as a function of $\theta$ for the different values of time $t$, with $\bar{n}_1=\bar{n}_2= 0,5$ is the mean number of a photon of the Gaussian environment,  and the initial squeezing parameter taking the value  $r=1,2$. This value is greater than the one used for GIP. This is due to the fact that EoF is not able to capture the existence of a quantum correlation when $r$ takes small values. This issue is discussed in Fig. (\ref{Fig. 3}). In our plots, we take the overall damping rate parameter at $\Gamma = 1$.}\label{Fig. 2}
\end{figure}

In Fig. (\ref{Fig. 2}), we depict the temporal behavior of the evolution GIP and EoF versus the initial rotation angle $\theta$ for various values of the time interaction with the Gaussian environment $t$. From the results reported in this figure, it is clear that the maximum amount of GIP and EoF is obtained when $\theta=\frac{\pi }{2}\left( {1 + k} \right)\hspace{0.3cm} \text{with} \hspace{0.2cm}k \in \mathbb{Z}$,  and for small values of time $t$. In addition,  the two behavior are similar and changes periodically as the periodic functions of period $\theta=\pi/2$. This indicates that the output state contains more amount of quantum correlations, especially for lower values of time $t$ and when $\theta=\frac{\pi }{2}\left( {1 + k} \right)$. We notice that the increase in time interaction with the Gaussian environment tends, in fact, to reduce the quantum correlations in the system. We also note that the results obtained by using GIP are plotted when the initial squeezing parameter is $r = 0,3$. Whereas those obtained by EoF are plotted when the initial squeezing parameter is $r = 1,2$. To clarify this difference, we will plot, in Fig. (\ref{Fig. 3}), the two behavior of GIP and EoF as the function of the initial squeezing parameter $r$.

\begin{figure}[H]
	\centering
	\begin{subfigure}{.49\textwidth}
		\centering
		\includegraphics[width=8cm]{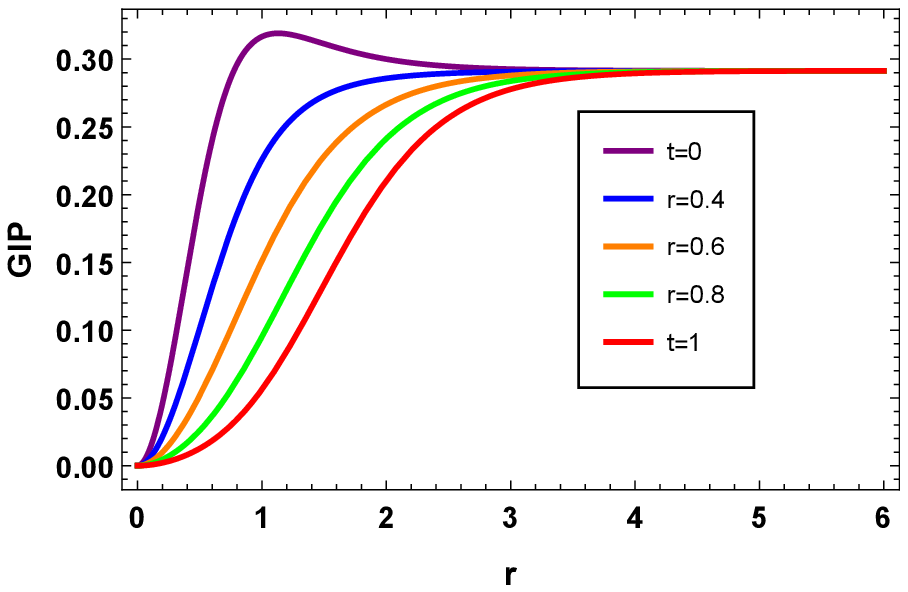}
		\caption{}\label{Fig. 3a}
	\end{subfigure}
	\begin{subfigure}{.49\textwidth}
		\centering
		\includegraphics[width=8cm]{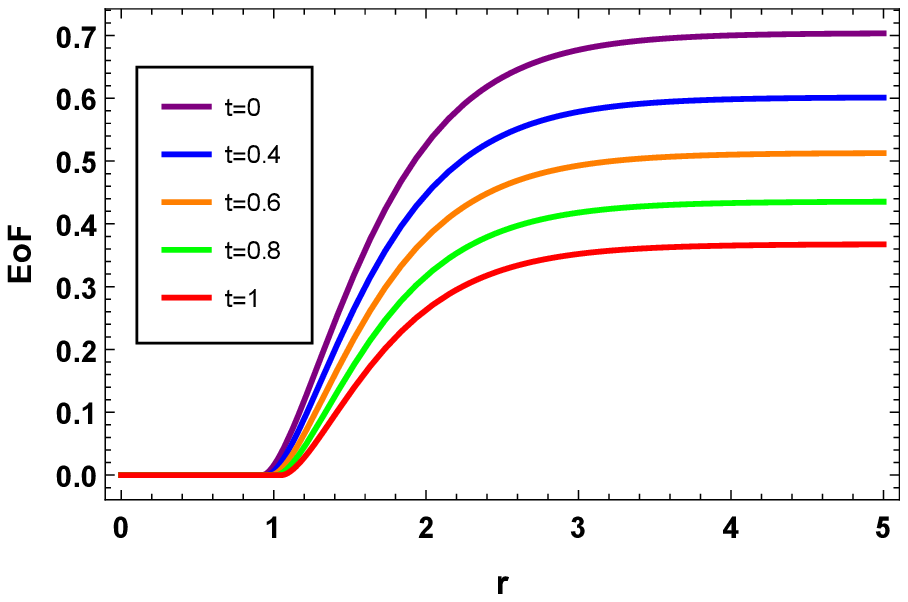}
		\caption{}\label{Fig. 3b}
	\end{subfigure}
	\renewcommand\thefigure{\arabic{figure}}
	\addtocounter{figure}{-1}
	\captionsetup{justification=raggedright, singlelinecheck=false, labelfont=sc} \captionof{figure}{The plot of Gaussian interferometric power and Gaussian entanglement of formation, for a two-mode output state, as a function of the initial squeezing parameter $r$. Fig. (\ref{Fig. 3a}) represents the behavior of GIP as a function of $r$ for the different values of time $t$, with $\bar{n}_1=\bar{n}_2= 0,5$ is the mean number of a photon of the Gaussian environment, and the initial rotation angle takes $\pi/2$. Fig. (\ref{Fig. 3b}) represents the behavior of EoF as a function of $r$ for the different values of time $t$, with $\bar{n}_1=\bar{n}_2= 0,5$ is the mean number of a photon of the Gaussian environment,  and the initial rotation angle $\pi/2$. In this plots, we take the overall damping rate parameter at $\Gamma = 1$.}\label{Fig. 3}
\end{figure}
Let us now analyze the behavior of quantum correlations measured by GIP beyond the quantum entanglement measured by EoF when we vary the initial squeezing parameter $r$ for various values of time $t$. As depicted in Fig. (\ref{Fig. 3a}) and Fig. (\ref{Fig. 3b}), it is obvious that the GIP and EoF have exhibited similar behavior as a function of $r$ for the various values of time $t$. Moreover,  in the limit when $r$ takes the smallest values, it is easy to see that the GIP can capture the presence of quantum correlations in comparison with EoF. Indeed, in the limit when $r$ takes values between 0 and 1, the EoF indicates the absence of quantum correlations, despite the condition that the two modes are correlated whenever the squeezing parameter is different from zero and in the limit of $r \to \infty$, we have an ideal entangled state with perfect correlations. One can interpret this by the weakness of EoF compared to GIP. That is clarified in the following Figure (Fig. (\ref{Fig. 4})), in which we will plot the temporal dynamic behavior of GIP and EoF as function of time interaction.
\begin{figure}[H]
	\centering
	\begin{subfigure}{.49\textwidth}
		\centering
		\includegraphics[width=8cm]{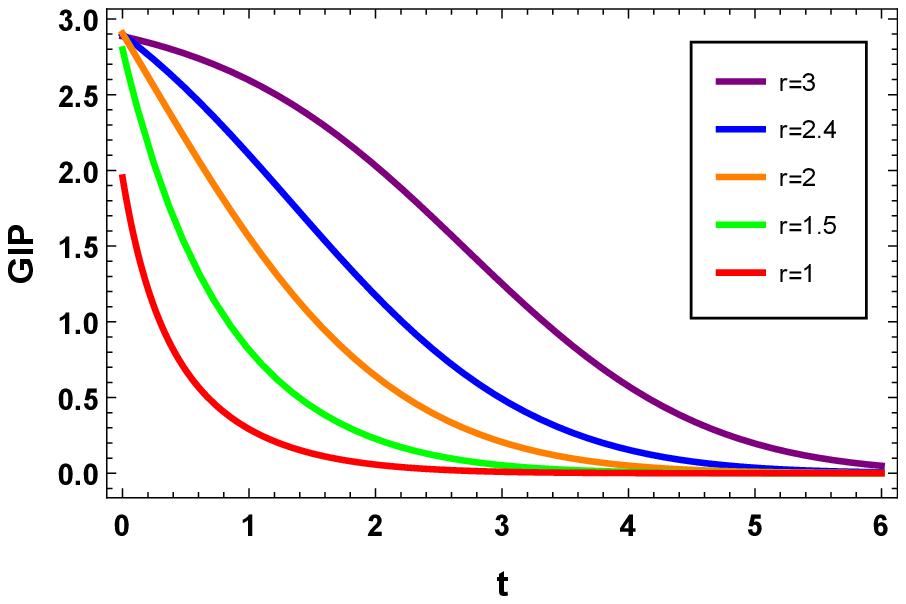}
		\caption{}\label{Fig. 4a}
	\end{subfigure}
	\begin{subfigure}{.49\textwidth}
		\centering
		\includegraphics[width=8cm]{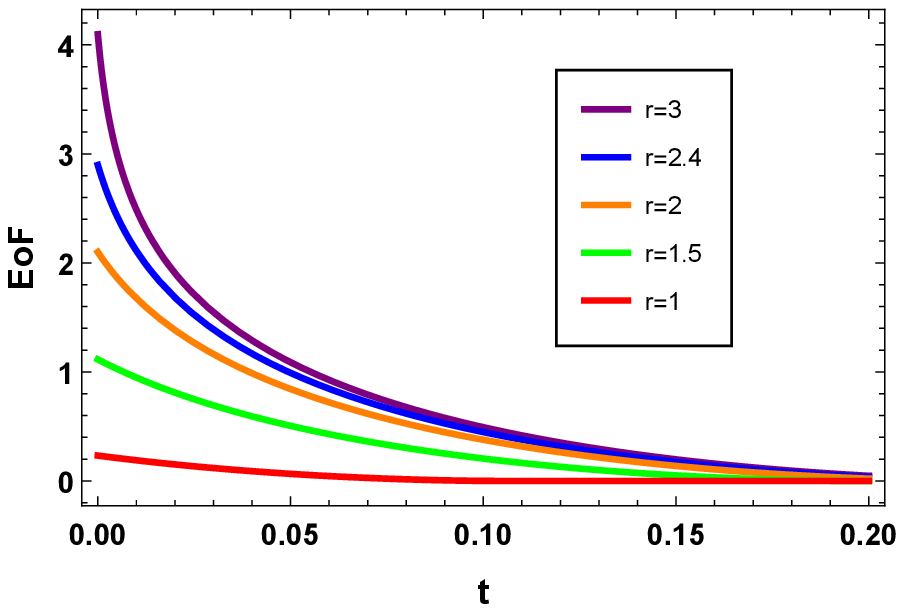}
		\caption{}\label{Fig. 4b}
	\end{subfigure}
	\renewcommand\thefigure{\arabic{figure}}
	\addtocounter{figure}{-1}
	\captionsetup{justification=raggedright, singlelinecheck=false, labelfont=sc} \captionof{figure}{The plot of Gaussian interferometric power and Gaussian entanglement of formation, for a two-mode output state, as functions of the interaction time $t$ with the Gaussian environment. Fig. (\ref{Fig. 4a}) represents the dynamic behavior of GIP as a function of time $t$ for the different values of the initial squeezing parameter $r$, with $\bar{n}_1=\bar{n}_2= 0,5$ is the mean number of a photon of the Gaussian environment, and the rotation angle is fixed as $\theta= \pi/2$. Fig. (\ref{Fig. 4b}) represents the dynamic behavior of EoF as a function of time $t$ for the different values of the initial squeezing parameter $r$ with $\bar{n}_1=\bar{n}_2= 0,5$ is the mean number of a photon of the Gaussian environment, and the rotation angle is fixed as $\theta= \pi/2$.  The overall damping rate parameter takes $\Gamma = 1$.}\label{Fig. 4}
\end{figure}

In Fig. (\ref{Fig. 4}), we depict the dynamics behavior of the GIP and EoF as a function of the interaction time with the Gaussian environment. From the results plotted in Fig. (\ref{Fig. 4a}) and Fig.  (\ref{Fig. 4a}), we notice that, in the limit of the period of interaction time $t$, the GIP and the EoF are able to qualify the degree of quantum correlations and quantum entanglement between the output modes. That is typically due to restriction of the effect of the Gaussian environment. While, in the increases of interaction time with the environment, the amount of correlation detected by EoF tends to zero. This fact is due to the influence imposed by the  Gaussian environment. Furthermore, the behavior dynamics temporal of the GIP shows its robustness against the effects of the Gaussian environment. This can be interpreted by the desired property that characterizes the GIP, which is a discord-type quantifier.

\section{Conclusion}

Quantum Gaussian states and their operations are at the heart of quantum information processing with continuous variables. The fundamental reason for this importance is that the vacuum state of quantum electrodynamics is itself a Gaussian state, which leads to its use in various quantum information processing tasks. Therefore, it is of special importance to examine and analyze the quantum correlations in Gaussian states, particularly quantum entanglement and quantum discord. In this work, we have presented a schema devoted to analyze the general non-classical correlations in continuous variable Gaussian noise channels, using bipartite Gaussian states as probes. We have described the quantum correlations of the two-mode Gaussian states under the Gaussian thermal environment by evaluating the  Gaussian interferometric power (GIP)  and  Gaussian entanglement of formation (EoF). The dynamics of these quantifiers were described in terms of the covariance matrix for Gaussian input states and the interaction time with the  Gaussian thermal environment. The comparison of these quantizers shows that GIP is able to capture a non-classical feature beyond the quantum entanglement that is evaluated by EoF. This non-classical feature is quantum correlation which reflects the robustness of GIP to withstand the noisy environment compared with EoF.
\section*{The data availability statement}
No additional data available

\end{document}